\documentclass[]{elsarticle}

\usepackage[english]{babel}
\usepackage{natbib}
\usepackage[colorlinks=true,linkcolor=black, citecolor=blue, urlcolor=blue]{hyperref}
\usepackage{graphicx}
\usepackage{placeins}
\usepackage{amsmath}
\usepackage{tikz}
\usetikzlibrary{arrows,chains,calc,shapes,automata,positioning,decorations.markings}
\usepackage{pgfplots}
\usepackage{color,soul}
\usepackage{lipsum}

\begin{document}

\begin{frontmatter}
	
\title{Deep Learning for One-dimensional Consolidation}
\author[]{Yared W. Bekele}
\address{{\scriptsize Rock and Soil Mechanics Group, SINTEF AS, Trondheim, Norway}}
\ead{yared.bekele@sintef.no}

\begin{abstract}
{\footnotesize Neural networks with physical governing equations as constraints have recently created a new trend in machine learning research. In line with such efforts, a deep learning model for one-dimensional consolidation where the governing equation is applied as a constraint in the neural network is presented here. A review of related research is first presented and discussed. The deep learning model relies on automatic differentiation for applying the governing equation as a constraint. The total loss is measured as a combination of the training loss (based on analytical and model predicted solutions) and the constraint loss (a requirement to satisfy the governing equation). Two classes of problems are considered: forward and inverse problems. The forward problems demonstrate the performance of a physically constrained neural network model in predicting solutions for one-dimensional consolidation problems. Inverse problems show prediction of the coefficient of consolidation. Terzaghi's problem with varying boundary conditions are used as example and the deep learning model shows a remarkable performance in both the forward and inverse problems. While the application demonstrated here is a simple one-dimensional consolidation problem, such a deep learning model integrated with a physical law has huge implications for use in, such as, faster real-time numerical prediction for digital twins, numerical model reproducibility and constitutive model parameter optimization.} 
\end{abstract}

\begin{keyword}
{\footnotesize deep learning \sep consolidation \sep forward problems \sep inverse problems}
\end{keyword}

\end{frontmatter}

\section{Introduction}

\noindent Machine learning has been, and still is, a fast growing field of research over the last years with ever growing areas of application outside pure computer science. Deep learning, a particular subset of machine learning which uses artificial neural networks, is being applied in various disciplines of science and engineering with usually surprising levels of success. Recently, the application of deep learning related to partial differential equations (PDEs) has been picking up pace. Several researchers are contributing to this effort where different names are given to the use of deep learning associated with physical systems governed by PDEs. Some of the commonly encountered labels include \emph{physics-informed neural networks}, \emph{physics-based deep learning}, \emph{theory-guided data science} and \emph{deep hidden physics models}, to name a few. In general, the aims of these applications include improving the efficiency, accuracy and generalization capability of numerical methods for the solution of PDEs. 

Data driven solution of partial differential equations was recently presented by \citet{raissi2019physics}. The authors investigated various differential equations and demonstrated how deep learning models can be applied in  a forward and inverse problem setting. Some of the PDEs studied include Burgers' and Navier-Stokes equations. The forward problems demonstrated how deep learning models can be trained based on sample data from exact solutions while optimizing an embedded governing partial differential equation. In the inverse problems, deep learning models were trained to identify the coefficients of the partial differential equations. For practical problems in science and engineering, this is equivalent to identifying material parameters based on given exact or numerical solutions. The neural network models for both the forward and inverse problems showed astonishing levels of accuracy. \citet{bar2019learning} presented a similar study where the emphasis was on learning data-driven discretizations that are best suited to a partial differential equation with certain boundary conditions. A related study combining deep learning and PDEs was presented by \citet{sirignano2018dgm} where the authors introduce a so-called Deep Galerkin Method (DGM). The proposed method was applied to different PDEs. This method was applied to application problems such as in quantitative finance and statistical mechanics, governed by the Black-Scholes and Focker-Planck PDEs, respectively; see \citet{al2018solving}. The application areas are increasing rapidly with different variations in the general methodology. A deep learning-based solution of the Euler equations for modeling high speed flows was presented by \citet{mao2020physics} where physics-informed neural networks were used for forward and inverse problems. Deep learning for computational fluid dynamics, in particular for vortex-induced vibrations, was presented by \citet{raissi2019deep}. A related work for predictive large-eddy-simulation wall modeling was presented by \citet{yang2019predictive}. The solution of time-dependent stochastic PDEs using physics-informed neural networks by learning in the modal space was demonstrated by \citet{zhang2019learning}. Application of deep learning, with physics-informed recurrent neural networks, to fleet prognosis was presented by \citet{nascimento2019fleet}. Bending analysis of Kirchhoff plates using a deep learning approach was shown by \citet{guo2019deep}. Deep learning-based study of linear and non-linear diffusion equations to learn parameters and unknown constitutive relationships was presented by \citet{tartakovsky2018learning}. Other recent and related studies are those by \citet{zhang2019quantifying}, \citet{yang2019adversarial}, \citet{meng2020composite}, \citet{sun2020surrogate}, \citet{huang2019predictive}, \citet{tipireddy2019comparative}, \citet{jia2020physics}, \citet{zheng2019physics} and \citet{xu2020physics}. While the neural network architecture used in many studies is a feed-forward network with the desired number of layers and hidden units, a recent study applied convolutional neural networks for the solution of the Poisson equation with varying meshes and Dirichlet boundary conditions; see \citet{ozbay2019poisson}. A conceptual framework for theory-guided deep learning is presented by \citet{karpatne2017theory}.

In this paper, application of deep learning to one-dimensional consolidation problems is presented. The problem describes fluid flow and excess pore water pressure dissipation in porous media. The governing equation for the problem is first discussed briefly. The deep learning model for the governing partial differential equation is then described. The problem is studied both for forward and inverse problems and the results from these are presented subsequently. Resources related to this work can be found on the author's GitHub page \href{https://github.com/yaredwb}{here}.

\FloatBarrier  

\section{Governing Equation}

The theory of consolidation describes the dissipation of fluid from a porous medium under compressive loading, thereby causing delay of the eventual deformation of the porous medium. The governing equation for one-dimensional consolidation is given by

\begin{equation}
\frac{\partial p}{\partial t} - \frac{\alpha m_v}{S + \alpha^2 m_v} \frac{\partial \sigma_{zz}}{\partial t} - c_v \frac{\partial^2 p}{\partial z^2} = 0
\label{eq:cons1D}
\end{equation}
where $p$ is the pore fluid pressure, $\alpha$ is Biot's coefficient, $S$ is the storativity of the pore space, $m_v$ is the confined compressibility of the porous medium, $ \sigma_{zz} $ is the the vertical effective stress and $c_v$ is the coefficient of consolidation. The classical one-dimensional consolidation problem is that at time $ t=0 $ a compressive load in the direction of fluid flow is applied and the load is maintained for $ t > 0 $; see \citet{verruijt2013theory}. This implies that, for $ t > 0 $, the stress $ \sigma_{zz} $ is constant, say with a magnitude $q$. Thus, we can reduce the general equation in \eqref{eq:cons1D} as

\begin{equation}
\begin{aligned}
\text{For} \; t = 0: \quad & p = p_o = \dfrac{\alpha m_v}{S + \alpha^2 m_v} q \\
\text{For} \; t > 0: \quad & \dfrac{\partial p}{\partial t} - c_v \dfrac{\partial^2 p}{\partial z^2} = 0
\end{aligned}
\label{eq:cons1Dt0andt}
\end{equation}

\begin{figure}[h]
	\centering
	\includegraphics[scale=0.5]{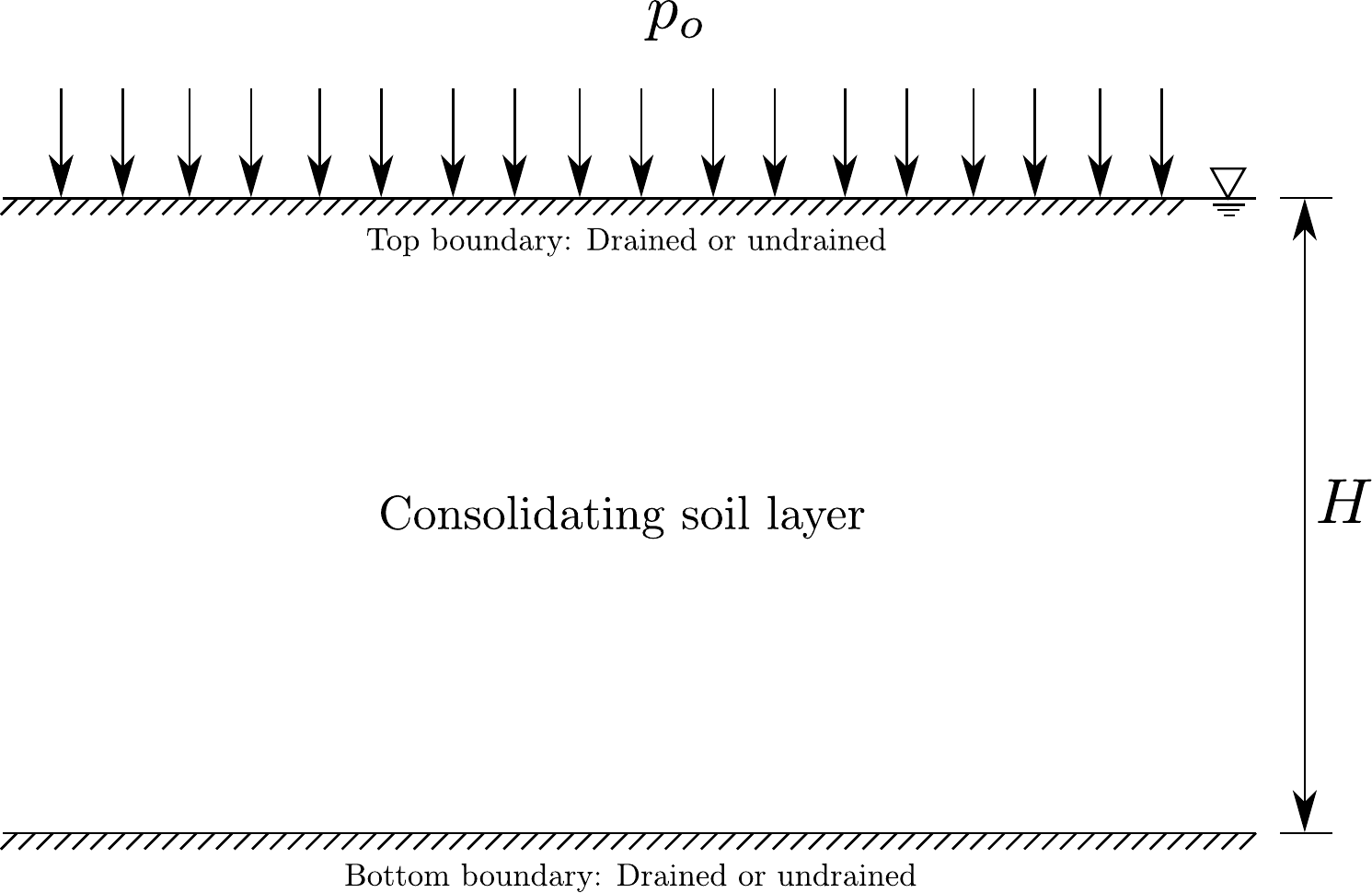}
	\caption{One-dimensional consolidation.}
\end{figure}

The first equation in \eqref{eq:cons1Dt0andt} establishes the initial condition for the 1D consolidation problem where we note that at $ t = 0 $ the total vertical load is carried by the pore fluid and we don't yet have any fluid dissipation from the porous medium. Whereas, the second equation governs the dissipation rate of the pore fluid as a function of both time and spatial dimension. This equation may be solved for various drainage boundary conditions at the top and bottom of the porous medium through either analytical or numerical methods. We consider an analytical solution here for two different drainage boundary conditions, which are described in a later section.   

\FloatBarrier

\section{Deep Learning Model}

In this section, the neural network architecture and the approach for a applying a physical constraint based on the governing one-dimensional consolidation equation are discussed. The model training procedure and the hyper-parameters that are controlled during training are also presented.

\subsection{Neural network architecture}

\begin{figure}[h]
	\centering
	\includegraphics[scale=0.42]{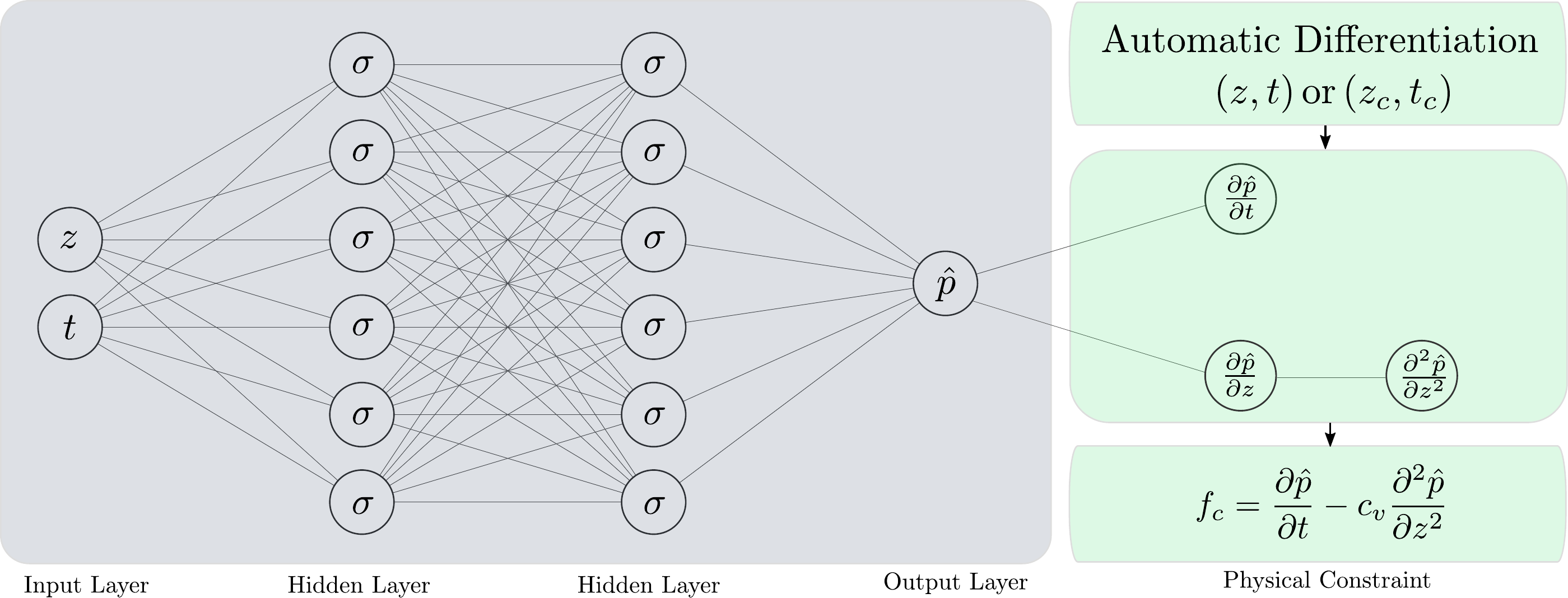}
	\caption{Illustration of the neural network architecture with input, hidden and output layers. The activation function used at the hidden units is $\sigma(x) = \tanh(x)$. Automatic differentiation is used to determined the partial derivatives in the governing equation and is used as a physical constraint to optimize together with the prediction error based on training data. The number of hidden layers and hidden units in this figure is for illustration only; the actual number of layers and hidden units used for different cases are discussed in a later section.}
	\label{fig:nn}
\end{figure}

A fully-connected deep neural network with the desired number of layers and hidden units is used as a model to be trained with the one-dimensional consolidation problem. An illustration of the neural network architecture is shown in Figure~\ref{fig:nn}. For the one-dimensional consolidation problem here, the input layer provides inputs of $(z,t)$ values from the training data, which usually includes initial and boundary condition data; the details are discussed in forward and inverse numerical example sections later.    
The neural network with the desired number of hidden layers and hidden units performs predictions of the excess pore pressure, which is then used to compute the loss based on the excess pore pressure training data. The neural network also includes a physical constraint based on the governing one-dimensional consolidation equation, where the constraint is evaluated using automatic differentiation, briefly discussed in a sub-section below. The neural network is designed to optimize both the \emph{training loss} and the physical constraint.

\subsection{Automatic differentiation}

A key part of the deep learning model for the problem here is automatic differentiation. It is important to not confuse automatic differentiation with other methods of computing derivatives in computer programs. There are four ways of computing derivatives using computers (\citet{baydin2017automatic}): a) manually obtaining the derivatives and coding them; b) numerical differentiation using finite difference approximations; c) computer-based symbolic differentiation and subsequent evaluation based on the algebraic expressions; and d) automatic differentiation, which is what is used here. Like the other methods, automatic differentiation provides numerical values of derivatives where these are obtained by using the rules of symbolic differentiation but by keeping track of derivative values instead of obtaining the final expressions. This approach of tracking derivative values makes automatic differentiation superior to the two most commonly used methods of computing derivatives, namely numerical differentiation and symbolic differentiation. Automatic differentiation exploits the fact that any derivative computation, no matter how complex, is composed of a sequence of elementary arithmetic operations and elementary function evaluations. It applies the chain rule repeatedly to these operations until the desired derivative is computed. Such an approach for computation makes automatic differentiation to be accurate at machine precision and computationally less demanding than other methods. For evaluating the derivatives in the governing one-dimensional equation here, the automatic differentiation capability in \verb|TensorFlow| is utilized. \verb|TensorFlow| provides an API for automatic differentiation by recording all operations and computing the gradients of the recorded computations using reverse mode differentiation; \citet{abadi2016tensorflow}. 

\subsection{Model training and hyper-parameters}

The deep learning model training is performed in slightly different ways for forward and inverse problems. However, the model hyper-parameters in both cases are adjusted in a similar way. For \emph{forward problems}, the training data involves initial and boundary condition data i.e. a pair of initial and boundary $(z,t)$ values and the corresponding excess pore pressure values $p(z,t)$. The model predicts the excess pore pressure value $\hat{p}$ for a given data point. The \emph{training loss} is calculated as a mean squared error from
\begin{equation}
MSE_p = \frac{1}{N} \sum_{k=1}^{N} \left| p(z_k,t_k) - \hat{p}(z_k,t_k) \right|^2,
\end{equation}   
where $N$ is the number of training data and $(z_k,t_k)$ represents the training data point. The physical constraint based on the governing partial differential equation is applied at randomly generated collocation points $(z_c,t_c)$. The collocation points are generated by using a latin hypercube sampling strategy where the bounds of the original training data are taken into consideration. The physical constraint is evaluated at the collocation points based on the predicted excess pore pressure using automatic differentiation i.e.
\begin{equation}
f_c = \frac{\partial \hat{p}}{\partial t_c} - c_v \frac{\partial^2 \hat{p}}{\partial z_c^2}.
\end{equation}
The \emph{constraint loss} is calculated as a mean squared error from
\begin{equation}
MSE_c = \frac{1}{N_c} \sum_{k=1}^{N_c} \left| f_c(z_{c,k},t_{c,k}) \right|^2,
\end{equation}
where $N_c$ is the number of collocation points and $(z_{c,k},t_{c,k})$ represents a data point from the collocation points. The total loss from training and the physical constraint is defined as
\begin{equation}
MSE = MSE_p + MSE_c,
\label{eq:total_loss}
\end{equation}
which is minimized by the model optimizer. For \emph{inverse problems}, the training procedure and loss evaluation is mostly similar with some differences. A larger size $(z,t)$ training data is used and collocation points are not generated in this case. This implies that automatic differentiation for the physical constraint is evaluated at the original training data points and the coefficient of consolidation is defined as a trainable parameter i.e.
\begin{equation}
f_c = \frac{\partial \hat{p}}{\partial t} - c_{vt} \frac{\partial^2 \hat{p}}{\partial z^2},
\end{equation}
where $c_{vt}$ is the trained value of the coefficient of consolidation updated during each step. An additional trainable model variable or weight $w_{cv}$, associated with the coefficient of consolidation, is introduced in the neural network and this is used to evaluate $c_{vt}$ based on an exponential function to always guarantee positive values during training i.e.
\begin{equation}
c_{vt} = \exp(w_{cv}).
\end{equation} 

The hyper-parameters tuned during training, both for forward and inverse problems include the \emph{number of layers}, \emph{number of hidden units}, \emph{batch size} and \emph{learning rate}. The batch size is adjusted to control the number of samples from the training data that are passed into the model before updating the trainable model parameters. The total loss here is minimized using the \emph{Adam optimizer} where its associated learning rate is tuned during the training process.

\FloatBarrier

\section{Forward Problems}
\label{sec:forward_problems}

\begin{figure}[h]
	\begin{center}
		\begin{tikzpicture}[scale=0.75]		
		\draw[fill=gray] (0,0) -- (1,0) -- (1,5) -- (0,5) -- (0,0);
		\draw[<->,>=stealth] (-0.5,0) -- (-0.5,5);
		\draw (-0.4,0) -- (-0.6,0);
		\draw (-0.4,5) -- (-0.6,5);
		\node [left] at (-1,2.5) {$ H $};
		\foreach \a [evaluate=\a as \a] in {0,0.2,0.4,0.6,0.8,1} {
			\draw[->,-stealth] (\a,5.5) -- (\a,5);
		}
		\draw (0,5.5) -- (1,5.5);
		\node [above] at (0.5,5.5) {$ p_o $};
		\node [below] at (0.5,5) {\small $ \Gamma_{t} $};
		\node [above] at (0.5,0) {\small $ \Gamma_{b} $};			
		\end{tikzpicture}
	\end{center}
	\caption{Example for one-dimensional consolidation.}
	\label{fig:1D_example}
\end{figure}
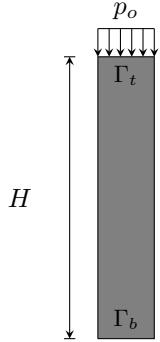

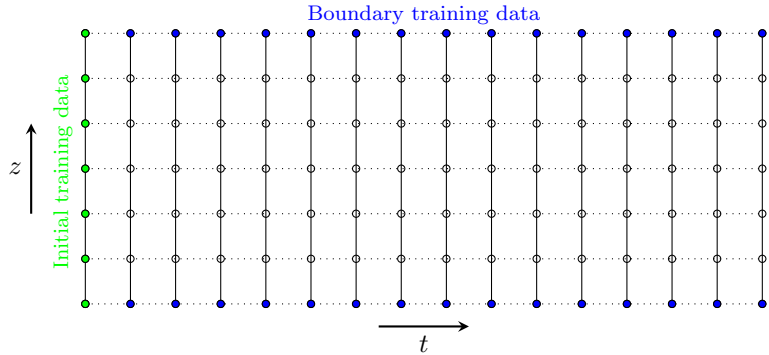
\begin{figure}[h]
	\begin{center}
		\begin{tikzpicture}[scale=1.2]		
		\foreach \z in {0, 0.5, 1, 1.5, 2, 2.5, 3}
		\foreach \t in {0, 0.5, 1, 1.5, 2, 2.5, 3, 3.5, 4, 4.5, 5, 5.5, 6.0, 6.5, 7.0, 7.5} {
			\draw [fill=white] (\t,\z) circle (0.04); 
		}
		\foreach \t in {0, 0.5, 1, 1.5, 2, 2.5, 3, 3.5, 4, 4.5, 5, 5.5, 6.0, 6.5, 7.0, 7.5} {
			\draw (\t,0) -- (\t,3);			
		}
		\foreach \z in {0, 0.5, 1, 1.5, 2, 2.5, 3} {
			\draw [dotted] (0,\z) -- (7.5,\z);
		}
		\draw[->,-stealth,thick] (-0.6,1) -- (-0.6,2);
		\draw[->,-stealth,thick] (3.25,-0.25) -- (4.25,-0.25);
		\node [left] at (-0.6,1.5) {$ z $};
		\node [below] at (3.75,-0.25) {$ t $};
		\foreach \t in {0, 0.5, 1, 1.5, 2, 2.5, 3, 3.5, 4, 4.5, 5, 5.5, 6.0, 6.5, 7.0, 7.5} {
			\draw [fill=blue] (\t,0) circle (0.04);
			\draw [fill=blue] (\t,3) circle (0.04); 
		}
		\foreach \z in {0, 0.5, 1, 1.5, 2, 2.5, 3} {
			\draw [fill=green] (0,\z) circle (0.04);
		}
		\node [text=blue, above] at (3.75,3) {\footnotesize Boundary training data};
		\node [text=green, rotate=90, left] at (-0.25,2.65) {\footnotesize Initial training data};
		\end{tikzpicture}
		\caption{Initial and boundary condition training data for forward problems. The total number $N$ of training data points $(z,t)$ depends on the spatial and temporal discretization used to obtain the exact solution. The training data is shuffled and divided into batches according a specified \emph{batch size} during training.}
		\label{fig:training_data_forward}
	\end{center}	
\end{figure}

The problem we will consider here is a classical one-dimensional consolidation problem, usually referred to as Terzaghi's problem, illustrated in Figure~\ref{fig:1D_example}. The problem examines the dissipation of excess pore pressure with time from the soil column due to the application of a surcharge load of magnitude $ p_o $ at the top boundary. This numerical example is studied for two drainage boundary conditions: with only the top boundary drained  and with both the top and bottom boundaries drained.
The training data for forward problems is selected to include initial and boundary condition data, as shown in Figure~\ref{fig:training_data_forward}. The nodes shown in the figure are for illustration only and the actual number of nodes and time steps depends on the spatial and temporal discretization used to obtain the analytical solution to the problem.
The total number of training points is divided into batches during training by using a specified batch size. The results from this numerical example for the two different boundary conditions are presented in the following sub-sections.

\FloatBarrier

\subsection{Consolidation with a drained top boundary}

The first case we will consider is a variation of Terzaghi's problem where excess pore pressure dissipation is allowed at the top boundary only i.e. the bottom boundary is considered impermeable. These boundary conditions are expressed mathematically as:

\begin{equation}
\begin{cases}
p = 0 \qquad & \text{at} \; \Gamma_{t}, \; t > 0 \\
\dfrac{\partial p}{\partial z} = 0 & \text{at} \; \Gamma_{b}, \; t > 0.
\end{cases}
\end{equation}
The initial condition is $ p=p_o $  for $ t=0 $. The analytical solution for the excess pore pressure as a ratio of the initial value is given by

\begin{equation}
\frac{p}{p_o} = \frac{4}{\pi} \sum_{k=1}^{\infty} \frac{(-1)^{k-1}}{2k-1} \cos\left[ (2k-1) \frac{\pi z}{2h} \right] \exp\left[ -(2k-1)^2 \frac{\pi^2}{4} \frac{c_v t}{h^2} \right] 
\label{eq:anasol}
\end{equation}
for $ t> 0 $, where $ h $ is the so-called \emph{drainage path} which in this case is equal to the total height of the domain i.e. $ h=H $. The spatial coordinate $z$ can be chosen to have its origin either at the top or at the bottom with positive coordinates in the domain. For the boundary conditions in this case we have $ 0 \leq z \leq H $.

For a numerical example, let's consider the height of the one-dimensional domain to be $ H = 1~\mathrm{m} $ and the coefficient of consolidation of the soil as $ c_v = 0.6~\mathrm{m}^2/\mathrm{yr} $. The exact solution based on the analytical solution is obtained by using a spatial discretization with $ N_z = 100 $ and a temporal discretization with $ N_t = 100 $, with $ t = 1 $ year. As illustrated in Figure~\ref{fig:training_data_forward}, the initial and boundary data $ \left\lbrace  z,t,p \right\rbrace $ are extracted from the exact solution as training data. The inputs to the neural network are the values $(z,t)$ where the model predicts a pore pressure value $\hat{p}$ as on output, which is then used to calculate the \emph{training loss}. A neural network with 10 hidden layers and 20 hidden units at each layer is used as the model to be trained. The spatial and temporal derivatives of the predicted pore pressure, as they appear in the constraint equation, are determined at selected collocation points using automatic differentiation in \verb|TensorFlow|. The collocation points are generated using the latin hypercube sampling strategy and for the example here $ N_c=10000 $ collocation $(z_c,t_c)$ points are generated. After the derivatives of $\hat{p}$ with respect to the collocation points are determined, the \emph{constraint loss} is calculated. A combination of training and constraint losses, as defined in equation~\eqref{eq:total_loss}, is minimized using the Adam optimizer for the desired number of epochs. The learning rate for the optimizer and the batch size used for training are 0.001 and 100, respectively.

\begin{figure}[h]
	\centering
	\includegraphics[scale=0.45]{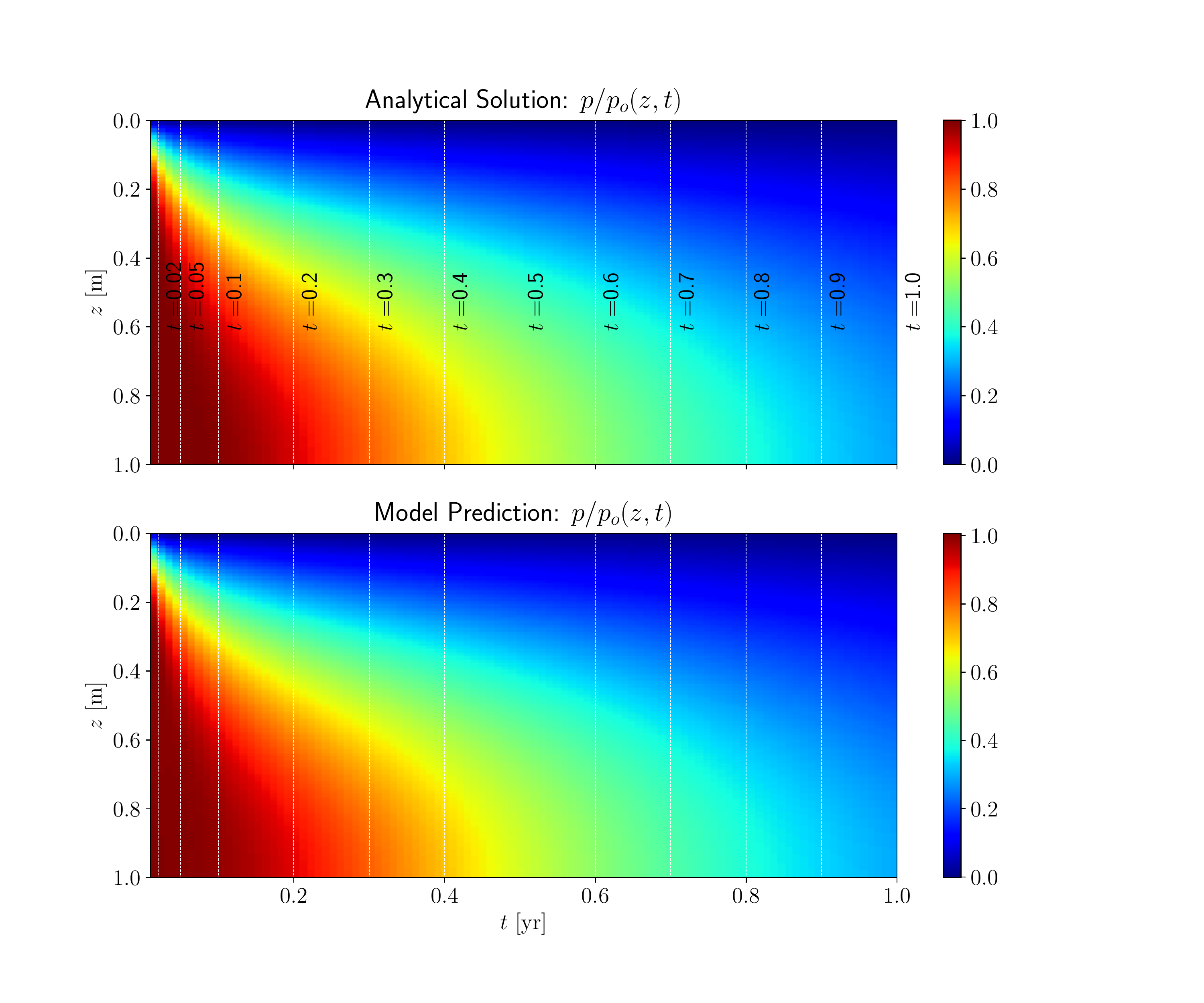}
	\caption{Results from analytical solution and model prediction in terms of color plots on $(z,t)$ grid for a drained top boundary. The color plots are obtained using interpolation with the nearest available values. The batch size used for training the model is 100.}
	\label{fig:color_plot_drained_top}
\end{figure}

The final model trained using the initial and boundary data and constrained at the collocation points according to the governing equation is used to predict the excess pore pressure for spatial and temporal points of the model domain. The results obtained from the analytical solution and model prediction are shown in Figure~\ref{fig:color_plot_drained_top} in terms of color plots on a two-dimensional grid from the $(z,t)$ data.
\begin{figure}[h]
	\centering
	\includegraphics[scale=0.42]{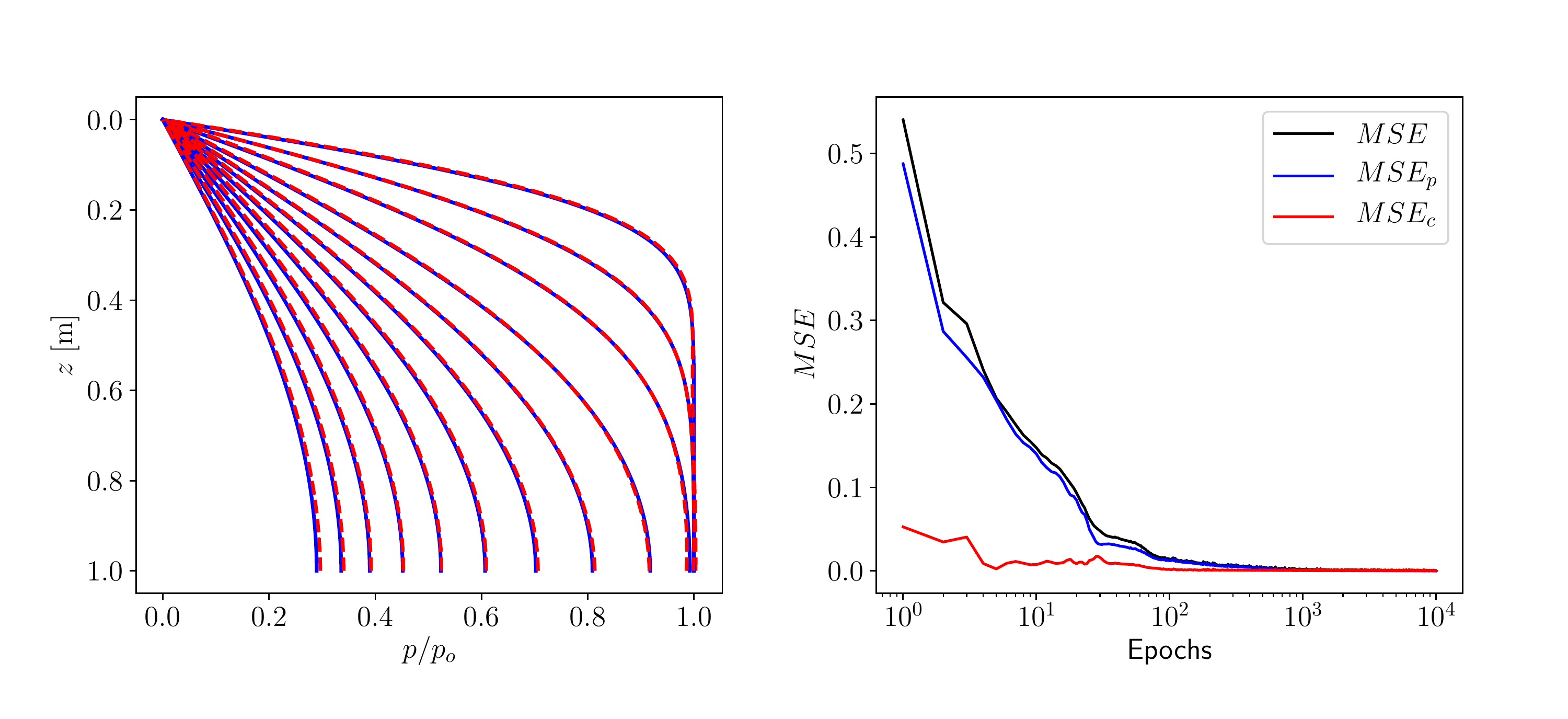}
	\caption{Left: Comparison of the excess pore pressure ratio between analytical solution (solid blue lines) and model prediction (red dashed lines) for selected time steps, for a drained top boundary. The time steps for comparison are those shown using vertical lines in the top plot in Figure~\ref{fig:color_plot_drained_top}. Right: Mean squared error versus number of epochs used for training.}
	\label{fig:plots_drained_top}
\end{figure}
The color plot is obtained by interpolation of the nearest excess pore pressure values from the actual grid points and is chosen here only for visualization convenience. As can be seen from the color plots of the analytical solution and model prediction, the deep learning model predicts the excess pore pressure values at the interior grid points reasonably well just based on initial and boundary training data. This demonstrates the remarkable accuracy of the physical constraint enforced through automatic differentiation in the deep learning model. A closer comparison of the analytical solution and model prediction is shown for selected time steps in the plot on the left in Figure~\ref{fig:plots_drained_top}. The time steps used for comparison are shown in the top plot in Figure~\ref{fig:color_plot_drained_top}. The results show a remarkably good agreement. The $L_2$ norm of the relative error i.e.
\begin{equation}
e = \frac{\| \hat{p} - p \|_{L_2}}{\| p \|_{L_2}},
\end{equation}
in this case was found to be $7.3\times10^{-3}$. A plot of the mean squared errors versus number of epochs is shown in the right plot in Figure~\ref{fig:plots_drained_top}. The plot shows the total mean squared error as well as the mean squared errors of the training and constraint losses. Mean squared error values in the order of $10^{-5}$ are obtained near the end of the training.    

\FloatBarrier

\subsection{Consolidation with drained top and bottom boundaries}

When both the top and bottom boundaries are permeable, excess pore pressure can dissipate through both boundaries. Mathematically, this boundary condition is expressed as

\begin{equation}
p = 0 \qquad  \text{for} \; \Gamma_{t} \cup \Gamma_{b}, \; t > 0 \\
\end{equation}
The analytical solution in equation \eqref{eq:anasol} still holds where in this case the drainage path is half of the height of the sample, i.e. $ h=H/2 $, as the pore fluid is allowed to dissipate through both the top and bottom boundaries. This case is equivalent to stating that there is no pore fluid flow at the center of the sample. Thus, the origin of the spatial coordinate is defined at the mid height of the domain and we have $ -H/2 \leq z \leq H/2 $.

The numerical example in the previous section is considered here again with the same model geometry but with different boundary conditions i.e. with drained top and bottom boundaries. In addition, the coefficient of consolidation in this case is assumed to be $ c_v = 0.1~\mathrm{m^2/yr} $ for the same model height, considering the faster consolidation as a result of both boundaries being drained. The analytical solution is again obtained using $ N_z = 100 $ and $ N_t = 100 $. The initial and boundary data are extracted in a similar way as in the previous case for training the model. The hyper-parameters of the neural network model are set to be similar as well. A deep network with 10 layers and 20 hidden units at each layer is used. The number of collocation points and the learning rate for the optimizer (Adam) are 10000 and 0.001, respectively. A batch size of 100 is used here as well.

\begin{figure}[h]
	\centering
	\includegraphics[scale=0.45, trim = 0cm 0.5cm 0cm 0.5cm, clip]{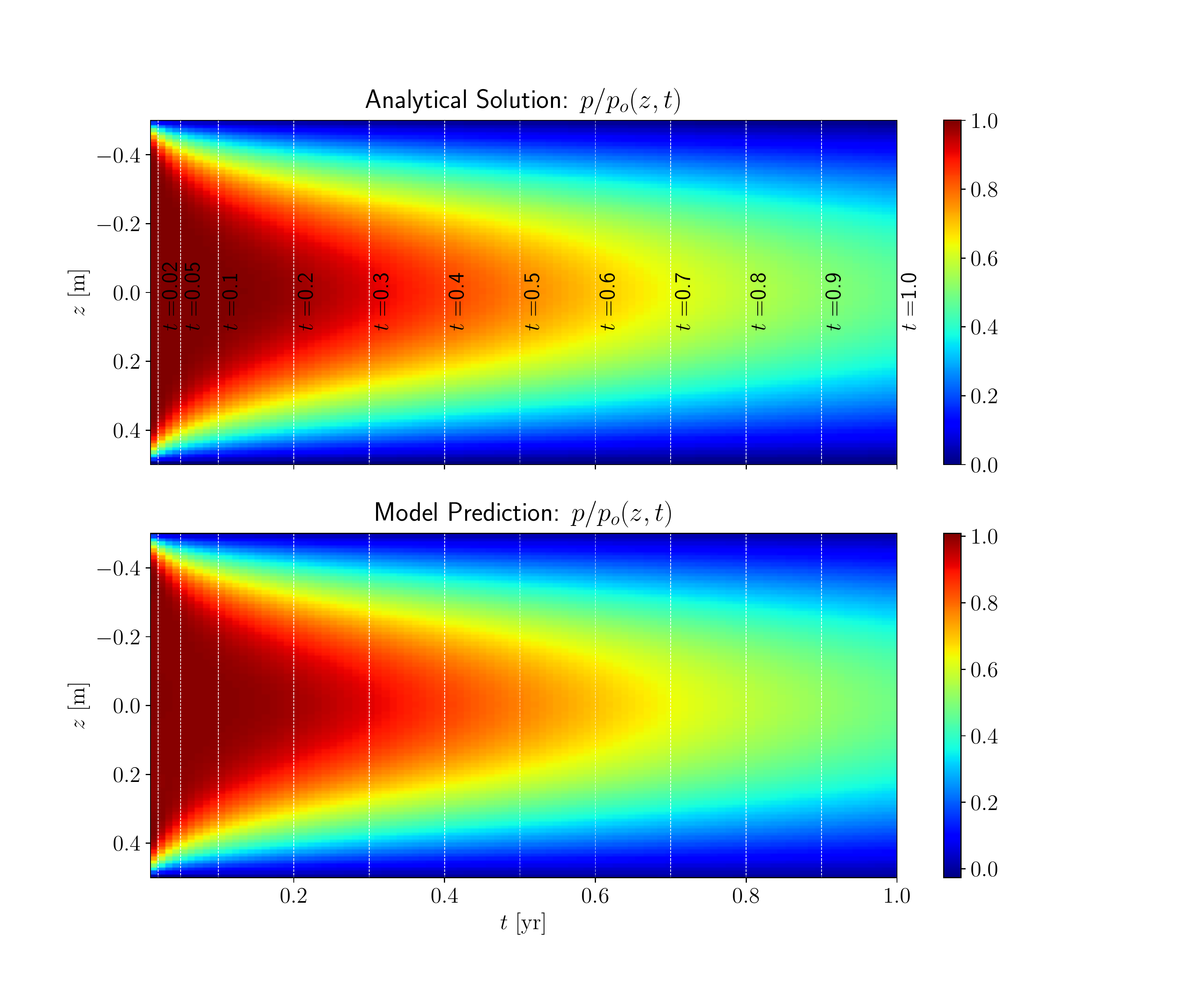}
	\caption{Results from analytical solution and model prediction in terms of color plots on $(z,t)$ grid for drained top and bottom boundaries. The color plots are obtained using interpolation with the nearest available values. The batch size used for training the model is 100.}
	\label{fig:color_plot_drained_top_and_bottom}
\end{figure}

Grid color plots comparing the analytical and model-predicted solutions are shown in Figure~\ref{fig:color_plot_drained_top_and_bottom}. We again observe a good performance by the deep learning model in predicting the excess pore pressure at the interior grid points.     
\begin{figure}[h]
	\centering
	\includegraphics[scale=0.42, trim = 0cm 0.5cm 0cm 0.5cm, clip]{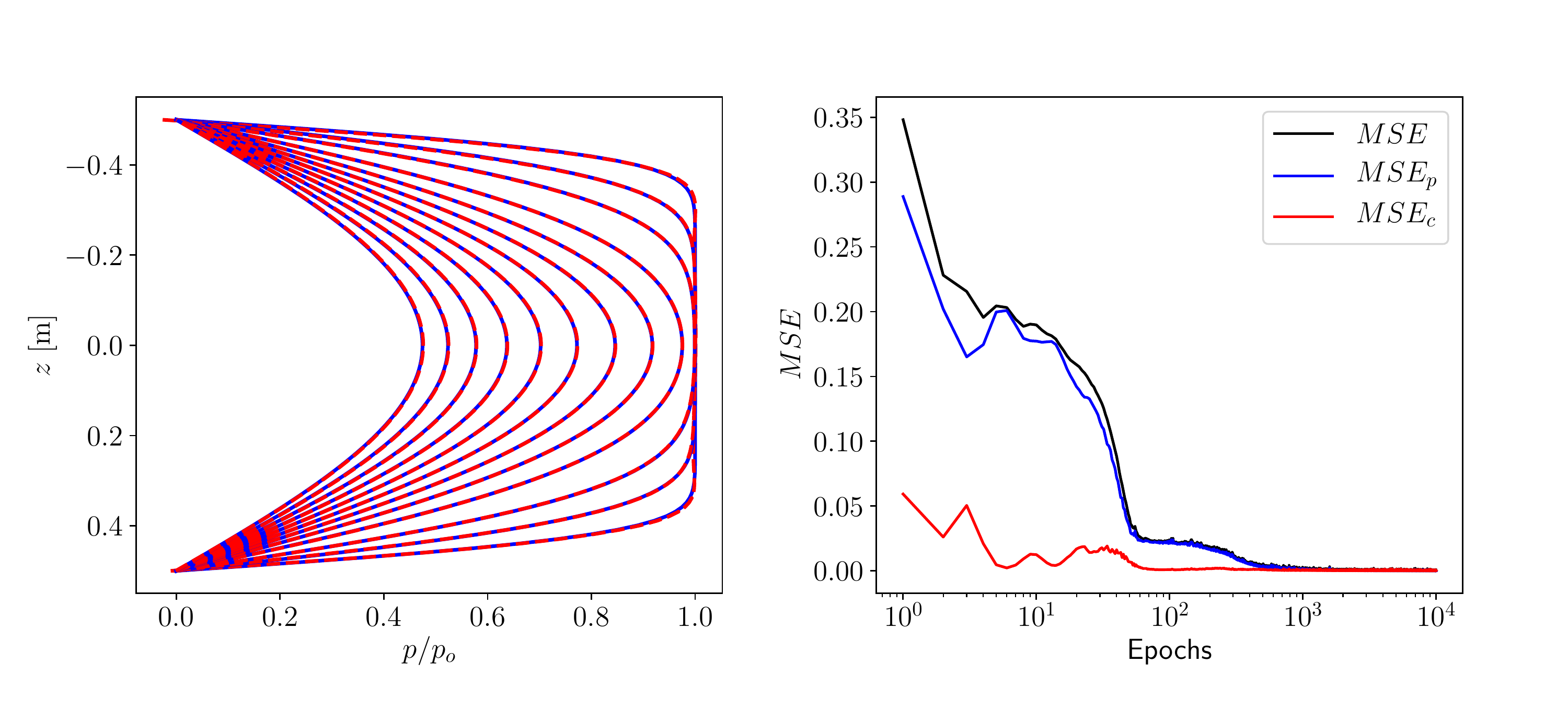}
	\caption{Left: Comparison of the excess pore pressure ratio between analytical solution (solid blue lines) and model prediction (red dashed lines) for selected time steps, for drained top and bottom boundaries. The time steps for comparison are those shown using vertical lines in the top plot in Figure~\ref{fig:color_plot_drained_top_and_bottom}. Right: Mean squared error versus number of epochs used for training.}
	\label{fig:plots_drained_top_and_bottom}
\end{figure}
A closer comparison of the excess pore pressure for selected time steps is shown in Figure~\ref{fig:plots_drained_top_and_bottom}. The time steps selected for comparison are shown in the top color plot in Figure~\ref{fig:color_plot_drained_top_and_bottom}. We see a very good agreement between the analytical solution and the deep learning model prediction. The $ L_2 $ norm of the relative error between the analytical and predicted solutions in this case is found to be  $ 2.62 \times 10^{-3} $. The second plot in Figure~\ref{fig:plots_drained_top_and_bottom} shows the evolution of the mean squared error with the number of epochs for the total mean squared error as well as the mean squared errors for the training and constraint losses. 

\FloatBarrier

\section{Inverse Problems}

\pgfmathdeclarerandomlist{MyRandomColors}{%
	{white}%
	{blue}%
}

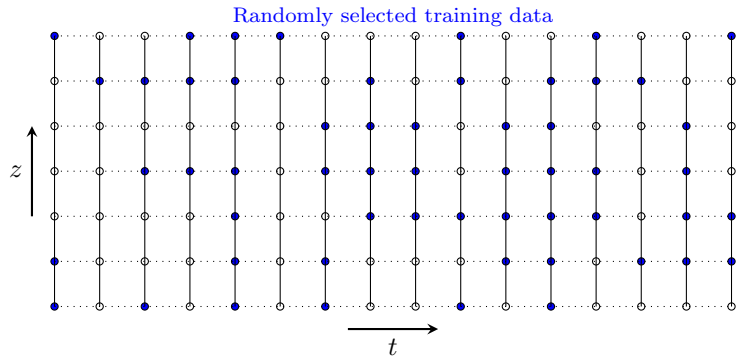
\begin{figure}[h]
	\begin{center}
		\begin{tikzpicture}[scale=1.2]
		\foreach \t in {0, 0.5, 1, 1.5, 2, 2.5, 3, 3.5, 4, 4.5, 5, 5.5, 6.0, 6.5, 7.0, 7.5}
		\foreach \z in {0, 0.5, 1, 1.5, 2, 2.5, 3}
		{
			\pgfmathrandomitem{\RandomColor}{MyRandomColors} 
			\draw[fill=\RandomColor] (\t,\z) circle (0.04);
		}
		\foreach \t in {0, 0.5, 1, 1.5, 2, 2.5, 3, 3.5, 4, 4.5, 5, 5.5, 6.0, 6.5, 7.0, 7.5} {
			\draw (\t,0) -- (\t,3);			
		}
		\foreach \z in {0, 0.5, 1, 1.5, 2, 2.5, 3} {
			\draw [dotted] (0,\z) -- (7.5,\z);
		}
		\draw[->,-stealth,thick] (-0.25,1) -- (-0.25,2);
		\draw[->,-stealth,thick] (3.25,-0.25) -- (4.25,-0.25);
		\node [left] at (-0.25,1.5) {$ z $};
		\node [below] at (3.75,-0.25) {$ t $};
		\node [text=blue, above] at (3.75,3) {\footnotesize Randomly selected training data};
		\end{tikzpicture}
		\caption{Randomly selected training data for inverse problems. The total number of training data points $(z,t)$ is chosen depending on the specific problem under consideration. The training data is shuffled and divided into batches according to the specified \emph{batch size} during training.}
		\label{fig:training_data_inverse}
	\end{center}
\end{figure}

The second class of problems we consider are inverse problems where we aim to find material/model parameters used in analysis based on a given solution to the problems. For our problem here, the aim is to determine the coefficient of consolidation used in an analytical solution given the excess pore pressure as a function of space and time i.e. $ p(z,t) $. The deep learning model is trained based on training data randomly selected from the whole analytical solution, as shown in Figure~\ref{fig:training_data_inverse}.
The size of the training data may be adjusted as desired but a limited sample size is usually sufficient for training to get good model prediction capabilities. Before training, the training data is shuffled and divided into batches based on a specified batch size. The main feature that differentiates inverse models from forward models, in the context of our problem here, is the fact that a trainable variable in addition to the default model variables is required to keep track of the coefficient of consolidation. Thus, a trainable weight is added to the deep learning model to update the coefficient of consolidation after initialization using a random weight. The constraint function based on the governing equation, with the trainable coefficient of consolidation, is continuously optimized together with the training loss. The following examples demonstrate this procedure numerically. It is worthwhile to mention here that even though we are dealing with a simple demonstration using a one-dimensional problem here, the approach has important implications in numerical modeling in science and engineering. Some of the potential applications of the methodology that could be mentioned are improved model reproducibility (given a certain numerical solution for a physical problem) and optimization of constitutive model parameters for complicated numerical simulations. These are important problems that may be addresses in a future work.

\FloatBarrier

\subsection{Consolidation with a drained top boundary}

The first example in the forward problems in Section \ref{sec:forward_problems} is considered here in an inverse setting with the same geometry and material/model parameters i.e. a one-dimensional model with a drained top boundary, a model height of $ H = 1~\mathrm{m} $ and a coefficient of consolidation of $ c_v = 0.6~\mathrm{m}^2/\mathrm{yr} $ is analyzed. The analytical solution is again obtained using $ N_z = 100 $ and $ N_t = 100 $. This implies that the number of $\left\lbrace z,t,p \right\rbrace $ points in the exact solution is equal to 10000. The neural network architecture is set up to have 10 hidden layers with 20 hidden units at each layer. A random sample of 2000 points is selected from the analytical solution data (with 10000 available points) for training the neural network. The training data is shuffled and divided into batches using a batch size of 200. The trainable weight corresponding to the coefficient of consolidation is initialized as $ w_{cv}=0 $, implying an initial coefficient of consolidation of $ 1.0~\mathrm{m}^2/\mathrm{yr} $. Adam optimizer is used to minimize the training and constraint losses with a learning rate of 0.0001.   

\begin{figure}[h]
	\centering
	\includegraphics[scale=0.45, trim = 0cm 1.5cm 0cm 1.5cm, clip]{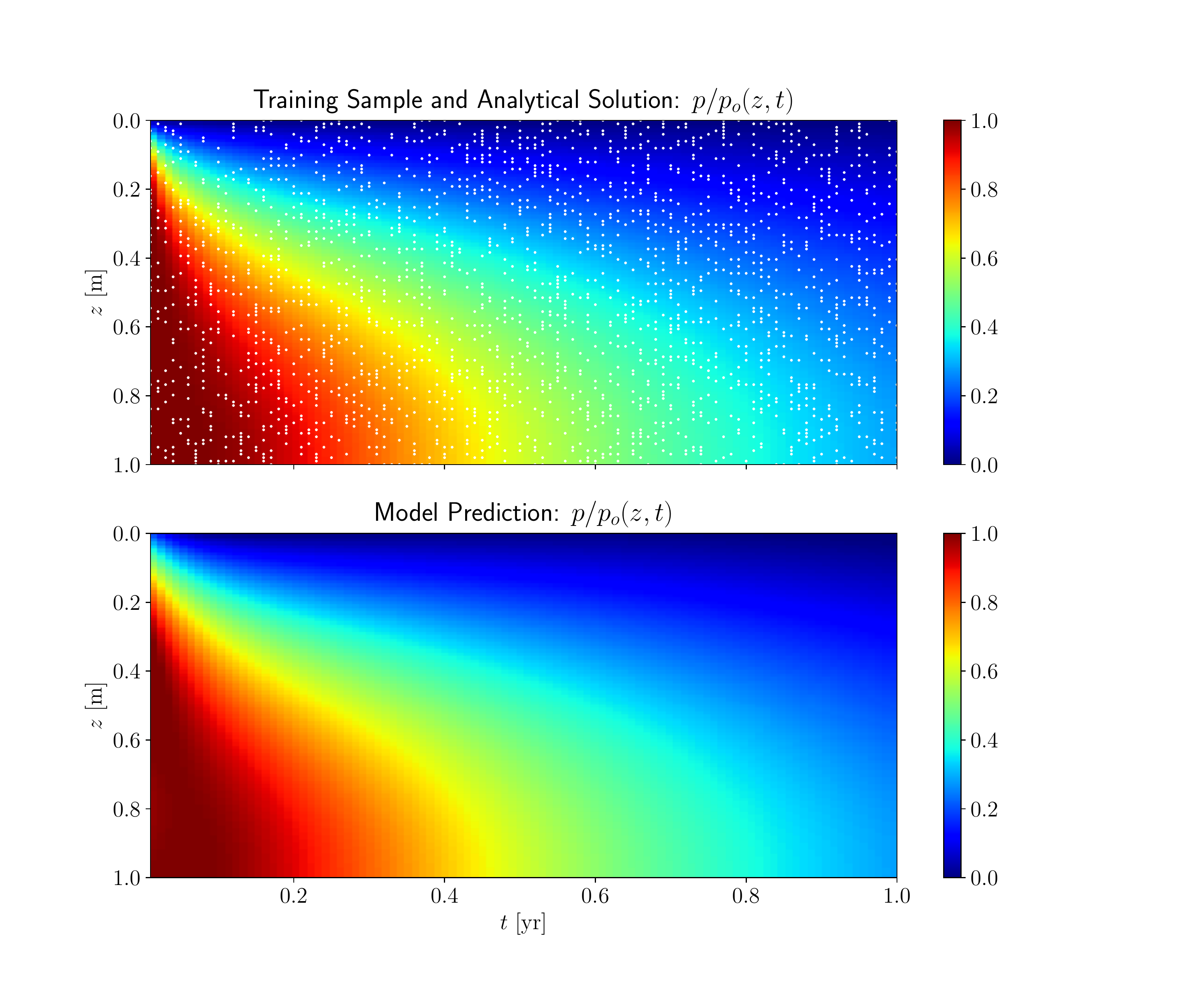}
	\caption{Inverse analysis results from analytical solution and model prediction in terms of color plots on $(z,t)$ grid for a drained top boundary. The color plots are obtained using interpolation with the nearest available values. The white dots represent the randomly selected training data points and the sample size used is 2000. The batch size used for training is 200.}
	\label{fig:color_plot_inverse_drained_top}
\end{figure}

The results for inverse analysis of the problem with a drained top boundary are shown in Figure~\ref{fig:color_plot_inverse_drained_top}. The randomly selected training data points are shown in the top color plot as white dots. As in the case of the forward problem, the deep learning model predicts the excess pore pressure well from a limited training sample data. This again shows the remarkable performance of the physical constraint obtained using automatic differentiation.   
\begin{figure}[h]
	\centering
	\includegraphics[scale=0.42]{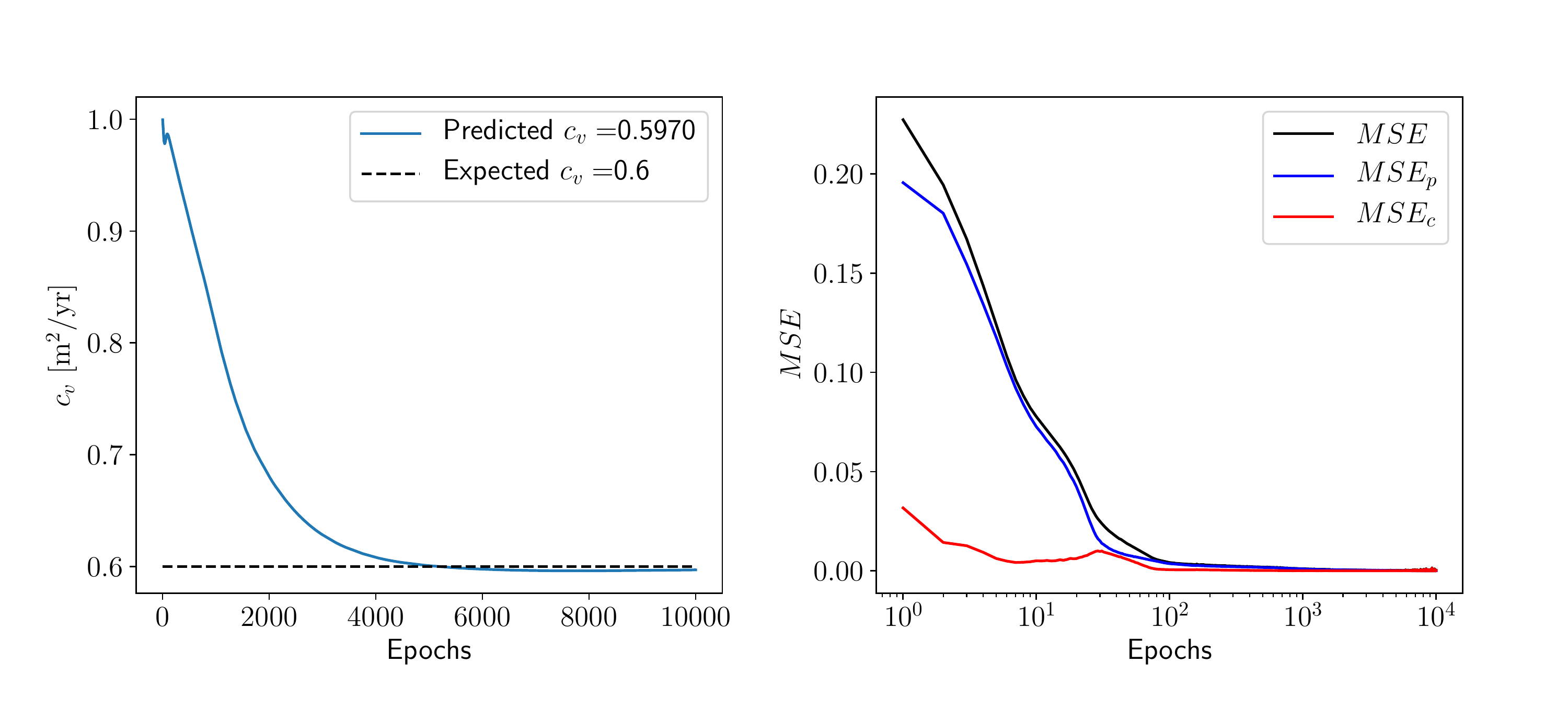}
	\vspace*{-10mm}
	\caption{Left: Evolution of the predicted coefficient of consolidation as a function of the number of training epochs, for a drained top boundary. Right: Mean squared error versus number of epochs used for training.}
	\label{fig:plots_inverse_drained_top}
\end{figure}
The plot on the left in Figure~\ref{fig:plots_inverse_drained_top} shows the evolution of the predicted coefficient of consolidation as the training progresses, as a function of the number of epochs. The final predicted value of the coefficient of consolidation is $ c_v = 0.5970~\mathrm{m^2/yr} $, which is close to the expected value of $ c_v = 0.6~\mathrm{m^2/yr} $ with an absolute error of $ 3.0 \times 10^{-3} $. The plot on the right in Figure~\ref{fig:plots_inverse_drained_top} shows the evolution of the mean squared errors as a function of the training epochs. 

\FloatBarrier

\subsection{Consolidation with drained top and bottom boundaries}

We consider again the numerical example in the previous section where in this case the top and bottom boundaries are drained. The model geometry remains similar but the analytical solution here is obtained using a coefficient of consolidation of $ c_v = 0.1~\mathrm{m^2/yr} $, which we aim to predict using the inverse deep learning model. We have 10000 data points from the analytical solution based on $ N_z = 100 $ and $ N_t = 100 $, from where a training sample of 2000 is randomly selected. The neural network has a similar architecture with 10 hidden layers and 20 hidden units at each layer. The other model hyper-parameters remain similar: the batch size is 200 and Adam optimizer is used with a learning rate of 0.0001.

\begin{figure}[h]
	\centering
	\includegraphics[scale=0.45, trim = 0cm 1.5cm 0cm 1.5cm, clip]{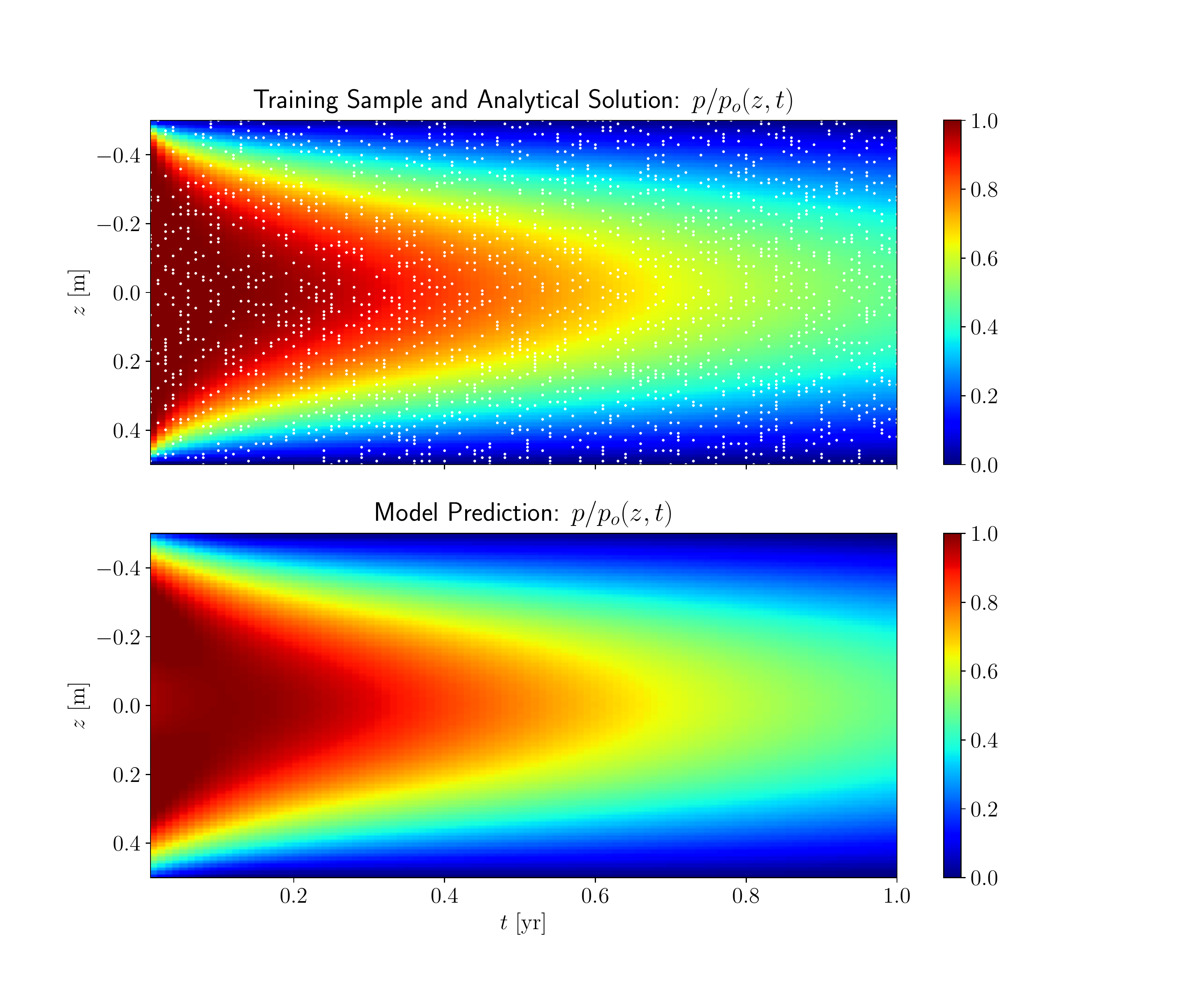}
	\caption{Inverse analysis results from analytical solution and model prediction in terms of color plots on $(z,t)$ grid for drained top and bottom boundaries. The color plots are obtained using interpolation with the nearest available values. The white dots represent the randomly selected training data points and the sample size used is 2000. The batch size used for training is 200.}
	\label{fig:color_plot_inverse_drained_top_and_bottom}
\end{figure}

The results for inverse analysis for drained top and bottom boundaries are shown in Figure~\ref{fig:color_plot_inverse_drained_top_and_bottom}. With a limited training sample, the model predicts the excess pore pressures throughout the grid with a good accuracy.  
\begin{figure}[h]
	\centering	
	\includegraphics[scale=0.42]{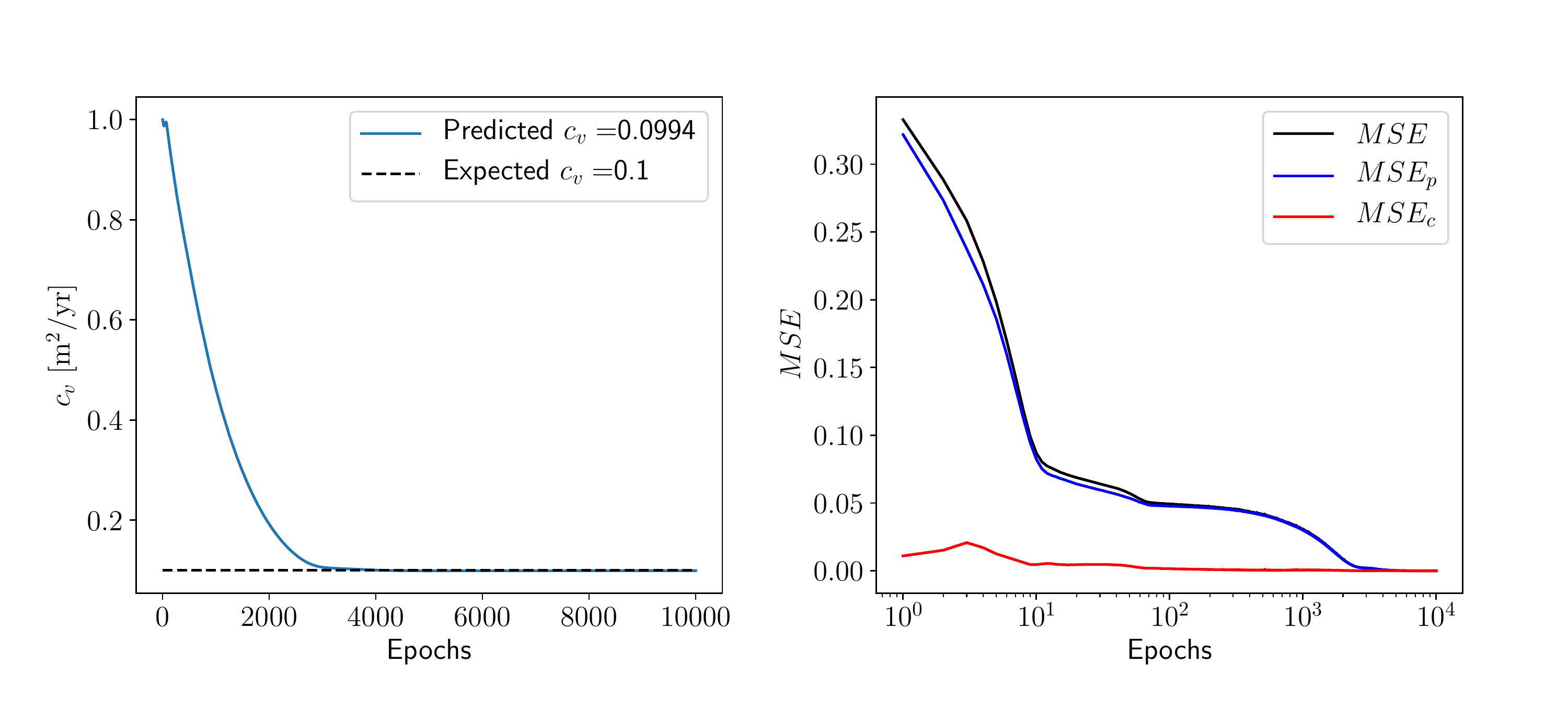}
	\vspace*{-10mm}
	\caption{Left: Evolution of the predicted coefficient of consolidation as a function of the number of training epochs, for drained top and bottom boundaries. Right: Mean squared error versus number of epochs used for training.}
	\label{fig:plots_inverse_drained_top_bottom}
\end{figure}
The coefficient of consolidation predicted by the deep learning model is $ c_v = 0.0994~\mathrm{m^2/yr} $, which compared to the expected values of $ c_v = 0.1~\mathrm{m^2/yr} $ implies an absolute error of $ 6.0 \times 10^{-4} $. These results are shown on the left plot in Figure~\ref{fig:plots_inverse_drained_top_bottom}, which shows the evolution of the predicted coefficient of consolidation as a function of the number of training epochs. The right plot in the same figure shows the mean squared errors (total, training and constraint) as a function of the number of training epochs.

\FloatBarrier

\section{Summary and Conclusions}

A deep learning model for one-dimensional consolidation is presented where the governing partial differential equation is used as a constraint in the model. Research on physics constrained neural networks has been gaining traction in recently in the machine learning research community and the work presented here adds to that effort. Various application areas where the idea has been applied were briefly reviewed, indicating the potential in diverse science and engineering disciplines.

The deep learning model used here is a fully-connected sequential neural network. The neural network is designed to take the spatial and temporal coordinates as inputs and predict the excess pore pressure, which is a function of these parameters. A key feature of the neural network architecture here is automatic differentiation which makes constraining of the system using the governing equation of one-dimensional consolidation. The spatial and temporal derivatives of the predicted excess pore pressure with respect to the spatial and temporal coordinates are evaluated using the automatic differentiation capability in \verb|TensorFlow|. Two classes of problems are considered: forward and inverse problems. For forward problems, the derivatives are evaluated at a certain number of collocation points which are generated using the latin hypercube sampling strategy, with the spatial and temporal bounds of the actual model taken into consideration. This approach made training of the neural using just the initial and boundary condition data possible with a remarkable degree of accuracy. For inverse problems, a larger size randomly selected data is used for training the neural network and the derivatives according to the governing equation are evaluated at the training data points. The coefficient of consolidation is used directly for forward problems while it is left as a trainable parameter for inverse problems. For both forward and inverse problems, the total training loss is defined as a combination of the training and constraint losses. The mean squared error for the training loss is evaluated based on the model predicted excess pore pressure and the corresponding exact analytical solution. The mean squared error for the constraint loss is evaluated according to the governing partial differential equation based on the derivatives evaluated using automatic differentiation. A model optimizer (Adam) is used to minimize the total mean squared error during training. For both classes of problems, the model hyper-parameters that are tuned include the number of hidden layers, number of hidden units at each layer, the batch size for training and the learning rate for the optimizer.  The potential of the model is demonstrated using Terzaghi's one-dimensional consolidation problem with two variations: with only a drained top boundary and with drained top and bottom boundaries. The model resulted in a remarkable prediction accuracy for both classes of problems; the pore pressure throughout the model's spatial and temporal bounds was predicted well for forward problems and the coefficient of consolidation was predicted with a very good accuracy for inverse problems.

While the application presented here is a simple one-dimensional consolidation problem, the implications of such a deep learning model are far greater. The efficiency demonstrated for forward problems indicates the potential for predicting numerical solutions faster using deep learning models trained using a very limited amount of data but with a very good accuracy because of the physical constraint. This could have a huge potential for digital twins where faster real time numerical prediction is desirable. The potential for predicting material parameters, as demonstrated using inverse problems, could be very useful in numerical model reproducibility and optimization of constitutive model parameters for complex numerical models.           

\bibliographystyle{elsarticle-num-names}
\bibliography{refs}

\begin{thebibliography}{25}
\expandafter\ifx\csname natexlab\endcsname\relax\def\natexlab#1{#1}\fi
\providecommand{\url}[1]{\texttt{#1}}
\providecommand{\href}[2]{#2}
\providecommand{\path}[1]{#1}
\providecommand{\DOIprefix}{doi:}
\providecommand{\ArXivprefix}{arXiv:}
\providecommand{\URLprefix}{URL: }
\providecommand{\Pubmedprefix}{pmid:}
\providecommand{\doi}[1]{\href{http://dx.doi.org/#1}{\path{#1}}}
\providecommand{\Pubmed}[1]{\href{pmid:#1}{\path{#1}}}
\providecommand{\bibinfo}[2]{#2}
\ifx\xfnm\relax \def\xfnm[#1]{\unskip,\space#1}\fi
\bibitem[{Raissi et~al.(2019)Raissi, Perdikaris, and
  Karniadakis}]{raissi2019physics}
\bibinfo{author}{M.~Raissi}, \bibinfo{author}{P.~Perdikaris},
  \bibinfo{author}{G.~E. Karniadakis},
\newblock \bibinfo{title}{Physics-informed neural networks: A deep learning
  framework for solving forward and inverse problems involving nonlinear
  partial differential equations},
\newblock \bibinfo{journal}{Journal of Computational Physics}
  \bibinfo{volume}{378} (\bibinfo{year}{2019}) \bibinfo{pages}{686--707}.
\bibitem[{Bar-Sinai et~al.(2019)Bar-Sinai, Hoyer, Hickey, and
  Brenner}]{bar2019learning}
\bibinfo{author}{Y.~Bar-Sinai}, \bibinfo{author}{S.~Hoyer},
  \bibinfo{author}{J.~Hickey}, \bibinfo{author}{M.~P. Brenner},
\newblock \bibinfo{title}{Learning data-driven discretizations for partial
  differential equations},
\newblock \bibinfo{journal}{Proceedings of the National Academy of Sciences}
  \bibinfo{volume}{116} (\bibinfo{year}{2019}) \bibinfo{pages}{15344--15349}.
\bibitem[{Sirignano and Spiliopoulos(2018)}]{sirignano2018dgm}
\bibinfo{author}{J.~Sirignano}, \bibinfo{author}{K.~Spiliopoulos},
\newblock \bibinfo{title}{Dgm: A deep learning algorithm for solving partial
  differential equations},
\newblock \bibinfo{journal}{Journal of Computational Physics}
  \bibinfo{volume}{375} (\bibinfo{year}{2018}) \bibinfo{pages}{1339--1364}.
\bibitem[{Al-Aradi et~al.(2018)Al-Aradi, Correia, Naiff, Jardim, and
  Saporito}]{al2018solving}
\bibinfo{author}{A.~Al-Aradi}, \bibinfo{author}{A.~Correia},
  \bibinfo{author}{D.~Naiff}, \bibinfo{author}{G.~Jardim},
  \bibinfo{author}{Y.~Saporito},
\newblock \bibinfo{title}{Solving nonlinear and high-dimensional partial
  differential equations via deep learning},
\newblock \bibinfo{journal}{arXiv preprint arXiv:1811.08782}
  (\bibinfo{year}{2018}).
\bibitem[{Mao et~al.(2020)Mao, Jagtap, and Karniadakis}]{mao2020physics}
\bibinfo{author}{Z.~Mao}, \bibinfo{author}{A.~D. Jagtap},
  \bibinfo{author}{G.~E. Karniadakis},
\newblock \bibinfo{title}{Physics-informed neural networks for high-speed
  flows},
\newblock \bibinfo{journal}{Computer Methods in Applied Mechanics and
  Engineering} \bibinfo{volume}{360} (\bibinfo{year}{2020})
  \bibinfo{pages}{112789}.
\bibitem[{Raissi et~al.(2019)Raissi, Wang, Triantafyllou, and
  Karniadakis}]{raissi2019deep}
\bibinfo{author}{M.~Raissi}, \bibinfo{author}{Z.~Wang}, \bibinfo{author}{M.~S.
  Triantafyllou}, \bibinfo{author}{G.~E. Karniadakis},
\newblock \bibinfo{title}{Deep learning of vortex-induced vibrations},
\newblock \bibinfo{journal}{Journal of Fluid Mechanics} \bibinfo{volume}{861}
  (\bibinfo{year}{2019}) \bibinfo{pages}{119--137}.
\bibitem[{Yang et~al.(2019)Yang, Zafar, Wang, and Xiao}]{yang2019predictive}
\bibinfo{author}{X.~Yang}, \bibinfo{author}{S.~Zafar}, \bibinfo{author}{J.-X.
  Wang}, \bibinfo{author}{H.~Xiao},
\newblock \bibinfo{title}{Predictive large-eddy-simulation wall modeling via
  physics-informed neural networks},
\newblock \bibinfo{journal}{Physical Review Fluids} \bibinfo{volume}{4}
  (\bibinfo{year}{2019}) \bibinfo{pages}{034602}.
\bibitem[{Zhang et~al.(2019)Zhang, Guo, and Karniadakis}]{zhang2019learning}
\bibinfo{author}{D.~Zhang}, \bibinfo{author}{L.~Guo}, \bibinfo{author}{G.~E.
  Karniadakis},
\newblock \bibinfo{title}{Learning in modal space: Solving time-dependent
  stochastic pdes using physics-informed neural networks},
\newblock \bibinfo{journal}{arXiv preprint arXiv:1905.01205}
  (\bibinfo{year}{2019}).
\bibitem[{Nascimento and Viana(2019)}]{nascimento2019fleet}
\bibinfo{author}{R.~G. Nascimento}, \bibinfo{author}{F.~A. Viana},
\newblock \bibinfo{title}{Fleet prognosis with physics-informed recurrent
  neural networks},
\newblock \bibinfo{journal}{arXiv preprint arXiv:1901.05512}
  (\bibinfo{year}{2019}).
\bibitem[{Guo et~al.(2019)Guo, Zhuang, and Rabczuk}]{guo2019deep}
\bibinfo{author}{H.~Guo}, \bibinfo{author}{X.~Zhuang},
  \bibinfo{author}{T.~Rabczuk},
\newblock \bibinfo{title}{A deep collocation method for the bending analysis of
  kirchhoff plate},
\newblock \bibinfo{journal}{Comput Mater Continua} \bibinfo{volume}{59}
  (\bibinfo{year}{2019}) \bibinfo{pages}{433--456}.
\bibitem[{Tartakovsky et~al.(2018)Tartakovsky, Marrero, Perdikaris,
  Tartakovsky, and Barajas-Solano}]{tartakovsky2018learning}
\bibinfo{author}{A.~M. Tartakovsky}, \bibinfo{author}{C.~O. Marrero},
  \bibinfo{author}{P.~Perdikaris}, \bibinfo{author}{G.~D. Tartakovsky},
  \bibinfo{author}{D.~Barajas-Solano},
\newblock \bibinfo{title}{Learning parameters and constitutive relationships
  with physics informed deep neural networks},
\newblock \bibinfo{journal}{arXiv preprint arXiv:1808.03398}
  (\bibinfo{year}{2018}).
\bibitem[{Zhang et~al.(2019)Zhang, Lu, Guo, and
  Karniadakis}]{zhang2019quantifying}
\bibinfo{author}{D.~Zhang}, \bibinfo{author}{L.~Lu}, \bibinfo{author}{L.~Guo},
  \bibinfo{author}{G.~E. Karniadakis},
\newblock \bibinfo{title}{Quantifying total uncertainty in physics-informed
  neural networks for solving forward and inverse stochastic problems},
\newblock \bibinfo{journal}{Journal of Computational Physics}
  \bibinfo{volume}{397} (\bibinfo{year}{2019}) \bibinfo{pages}{108850}.
\bibitem[{Yang and Perdikaris(2019)}]{yang2019adversarial}
\bibinfo{author}{Y.~Yang}, \bibinfo{author}{P.~Perdikaris},
\newblock \bibinfo{title}{Adversarial uncertainty quantification in
  physics-informed neural networks},
\newblock \bibinfo{journal}{Journal of Computational Physics}
  \bibinfo{volume}{394} (\bibinfo{year}{2019}) \bibinfo{pages}{136--152}.
\bibitem[{Meng and Karniadakis(2020)}]{meng2020composite}
\bibinfo{author}{X.~Meng}, \bibinfo{author}{G.~E. Karniadakis},
\newblock \bibinfo{title}{A composite neural network that learns from
  multi-fidelity data: Application to function approximation and inverse pde
  problems},
\newblock \bibinfo{journal}{Journal of Computational Physics}
  \bibinfo{volume}{401} (\bibinfo{year}{2020}) \bibinfo{pages}{109020}.
\bibitem[{Sun et~al.(2020)Sun, Gao, Pan, and Wang}]{sun2020surrogate}
\bibinfo{author}{L.~Sun}, \bibinfo{author}{H.~Gao}, \bibinfo{author}{S.~Pan},
  \bibinfo{author}{J.-X. Wang},
\newblock \bibinfo{title}{Surrogate modeling for fluid flows based on
  physics-constrained deep learning without simulation data},
\newblock \bibinfo{journal}{Computer Methods in Applied Mechanics and
  Engineering} \bibinfo{volume}{361} (\bibinfo{year}{2020})
  \bibinfo{pages}{112732}.
\bibitem[{Huang et~al.(2019)Huang, Xu, Farhat, and Darve}]{huang2019predictive}
\bibinfo{author}{D.~Z. Huang}, \bibinfo{author}{K.~Xu},
  \bibinfo{author}{C.~Farhat}, \bibinfo{author}{E.~Darve},
\newblock \bibinfo{title}{Predictive modeling with learned constitutive laws
  from indirect observations},
\newblock \bibinfo{journal}{arXiv preprint arXiv:1905.12530}
  (\bibinfo{year}{2019}).
\bibitem[{Tipireddy et~al.(2019)Tipireddy, Perdikaris, Stinis, and
  Tartakovsky}]{tipireddy2019comparative}
\bibinfo{author}{R.~Tipireddy}, \bibinfo{author}{P.~Perdikaris},
  \bibinfo{author}{P.~Stinis}, \bibinfo{author}{A.~Tartakovsky},
\newblock \bibinfo{title}{A comparative study of physics-informed neural
  network models for learning unknown dynamics and constitutive relations},
\newblock \bibinfo{journal}{arXiv preprint arXiv:1904.04058}
  (\bibinfo{year}{2019}).
\bibitem[{Jia et~al.(2020)Jia, Willard, Karpatne, Read, Zwart, Steinbach, and
  Kumar}]{jia2020physics}
\bibinfo{author}{X.~Jia}, \bibinfo{author}{J.~Willard},
  \bibinfo{author}{A.~Karpatne}, \bibinfo{author}{J.~S. Read},
  \bibinfo{author}{J.~A. Zwart}, \bibinfo{author}{M.~Steinbach},
  \bibinfo{author}{V.~Kumar},
\newblock \bibinfo{title}{Physics-guided machine learning for scientific
  discovery: An application in simulating lake temperature profiles},
\newblock \bibinfo{journal}{arXiv preprint arXiv:2001.11086}
  (\bibinfo{year}{2020}).
\bibitem[{Zheng et~al.(2019)Zheng, Zeng, and Karniadakis}]{zheng2019physics}
\bibinfo{author}{Q.~Zheng}, \bibinfo{author}{L.~Zeng}, \bibinfo{author}{G.~E.
  Karniadakis},
\newblock \bibinfo{title}{Physics-informed semantic inpainting: Application to
  geostatistical modeling},
\newblock \bibinfo{journal}{arXiv preprint arXiv:1909.09459}
  (\bibinfo{year}{2019}).
\bibitem[{Xu and Darve(2020)}]{xu2020physics}
\bibinfo{author}{K.~Xu}, \bibinfo{author}{E.~Darve},
\newblock \bibinfo{title}{Physics constrained learning for data-driven inverse
  modeling from sparse observations},
\newblock \bibinfo{journal}{arXiv preprint arXiv:2002.10521}
  (\bibinfo{year}{2020}).
\bibitem[{{\"O}zbay et~al.(2019){\"O}zbay, Laizet, Tzirakis, Rizos, and
  Schuller}]{ozbay2019poisson}
\bibinfo{author}{A.~G. {\"O}zbay}, \bibinfo{author}{S.~Laizet},
  \bibinfo{author}{P.~Tzirakis}, \bibinfo{author}{G.~Rizos},
  \bibinfo{author}{B.~Schuller},
\newblock \bibinfo{title}{Poisson cnn: Convolutional neural networks for the
  solution of the poisson equation with varying meshes and dirichlet boundary
  conditions},
\newblock \bibinfo{journal}{arXiv preprint arXiv:1910.08613}
  (\bibinfo{year}{2019}).
\bibitem[{Karpatne et~al.(2017)Karpatne, Atluri, Faghmous, Steinbach, Banerjee,
  Ganguly, Shekhar, Samatova, and Kumar}]{karpatne2017theory}
\bibinfo{author}{A.~Karpatne}, \bibinfo{author}{G.~Atluri},
  \bibinfo{author}{J.~H. Faghmous}, \bibinfo{author}{M.~Steinbach},
  \bibinfo{author}{A.~Banerjee}, \bibinfo{author}{A.~Ganguly},
  \bibinfo{author}{S.~Shekhar}, \bibinfo{author}{N.~Samatova},
  \bibinfo{author}{V.~Kumar},
\newblock \bibinfo{title}{Theory-guided data science: A new paradigm for
  scientific discovery from data},
\newblock \bibinfo{journal}{IEEE Transactions on Knowledge and Data
  Engineering} \bibinfo{volume}{29} (\bibinfo{year}{2017})
  \bibinfo{pages}{2318--2331}.
\bibitem[{Verruijt(2013)}]{verruijt2013theory}
\bibinfo{author}{A.~Verruijt},
\newblock \bibinfo{title}{Theory and problems of poroelasticity},
\newblock \bibinfo{journal}{Delft University of Technology}
  (\bibinfo{year}{2013}) \bibinfo{pages}{71}.
\bibitem[{Baydin et~al.(2017)Baydin, Pearlmutter, Radul, and
  Siskind}]{baydin2017automatic}
\bibinfo{author}{A.~G. Baydin}, \bibinfo{author}{B.~A. Pearlmutter},
  \bibinfo{author}{A.~A. Radul}, \bibinfo{author}{J.~M. Siskind},
\newblock \bibinfo{title}{Automatic differentiation in machine learning: a
  survey},
\newblock \bibinfo{journal}{The Journal of Machine Learning Research}
  \bibinfo{volume}{18} (\bibinfo{year}{2017}) \bibinfo{pages}{5595--5637}.
\bibitem[{Abadi et~al.(2016)Abadi, Agarwal, Barham, Brevdo, Chen, Citro,
  Corrado, Davis, Dean, Devin et~al.}]{abadi2016tensorflow}
\bibinfo{author}{M.~Abadi}, \bibinfo{author}{A.~Agarwal},
  \bibinfo{author}{P.~Barham}, \bibinfo{author}{E.~Brevdo},
  \bibinfo{author}{Z.~Chen}, \bibinfo{author}{C.~Citro}, \bibinfo{author}{G.~S.
  Corrado}, \bibinfo{author}{A.~Davis}, \bibinfo{author}{J.~Dean},
  \bibinfo{author}{M.~Devin}, et~al.,
\newblock \bibinfo{title}{Tensorflow: Large-scale machine learning on
  heterogeneous distributed systems},
\newblock \bibinfo{journal}{arXiv preprint arXiv:1603.04467}
  (\bibinfo{year}{2016}).

\end{thebibliography}

\end{document}